# Experimental Observation of Transitions of Different Pulse Solutions of Ginzburg-Landau Equation in a Mode-Locked Fiber Laser

**Junsong Peng, Li Zhan*, Zhaochang Gu, Shouyu Luo, and Qishun Shen**

*Department of Physics, Key Laboratory for Laser Plasmas (Ministry of Education), State Key Lab of Advanced Optical Communication Systems and Networks, Shanghai Jiao Tong University, Shanghai, 200240, China*

Transitions between different kinds of soliton solutions of Ginzburg-Landau equation (GLE) have been studied experimentally in a mode-locked fiber laser. It is demonstrated that the different kinds of solitons corresponding to different solutions of GLE can be generated in a single mode-locked laser. Dispersion-managed solitons (DM), all-normal-dispersion solitons (ANDi) and similaritons can be emitted respectively depending on the parameter of the intensity of the light field and the birefringence effect. The three nonlinear waves show different features especially the spectrum shapes and dynamics accompanying with pump power scaling. Such phenomenon reveals the properties of GLE, which is not only scientifically interesting but also valuable to practical applications of mode-locked fiber lasers.

PACS numbers: 42.55.Wd, 42.65.Re, 42.65.Tg, 42.65.Sf

Passively mode-locked fiber lasers have developed extensively since its first realization in 1990. It's not only valuable to the application in many fields, but also attractive to study the nonlinear phenomenon in physics such as different kinds of solitons. So far, there are mainly four types of solitons in mode-locked fiber lasers. They are nonlinear Schrödinger solitons (NSS) [1], DM solitons, ANDi and similaritons [2]. GLE can be used to describe these solitons in mode-locked lasers effectively. Usually, different kinds of solitons have to be generated in different laser cavities regarding the coefficients of GLE, especially the net dispersion. NSS exists in the cavities with anomalous dispersion, and DM solitons generate in the cavities consisting of segments of anomalous and normal dispersion. ANDi exists in all normal dispersion cavities. Recently, NSS and similaritons are found to coexist in just one laser cavity [3]. The pulse propagates self-similarly in the gain fiber with normal dispersion, and gradually evolves into a soliton in the rest of the cavity with anomalous dispersion. Also, it's found numerically that different soliton solutions coexist for the same set of parameters of GLE [4]. Particularly, numerical study implies that Gaussian soliton and similariton can be switched in a net positive dispersion cavity by adjusting the gain parameter [5]. Besides, transition between similariton and ANDi was observed numerically in a laser depending on the laser parameter [6]. It's thus intuitive to get different kinds of solitons in a fixed dispersion map i.e. a single laser. However, to the best of our knowledge, the corresponding experimental observation hasn't been discovered so far.

In the present work, we construct a laser to study GLE experimentally. Finally, different kinds of soliton solutions (DM solitons, ANDi and similaritons) can be observed respectively depending on the parameter such as the strength of the light field and the birefringence of the fiber. Mathematically, these different kinds of solitons correspond to the solutions to GLE under different conditions. Varying the coefficients of GLE gives various soliton solutions, corresponding to transitions between different solitons observed. DM solitons can be emitted when the light field is weak. ANDi is generated as the strength of the field is increased. Similaritons are found as long as the spectrum width of ANDi is decreased to certain value. The three solitons can be distinguished by different spectrum shape experimentally. Furthermore, the spectrum width of ANDi increases with pump power, which is consistent with the theoretical predictions by V.L. Kalashnikov [7] and A. Cabasse [8]. However, it's quite different for DM solitons. Haus theoretically shows that the spectrum width of DM decreases with gain increment [9]. We further check this for the demonstration of DM solitons, which hasn't been found experimentally to our knowledge.

Although transitions between different solitons were theoretically studied based on GLE, the corresponding experimental observations are lagged, since it's a challenge to relate the coefficients of GLE to the physical parameters in a laser system. Whether varying the coefficients in experiments is possible is still an open question. Our observation answers the question and links the coefficients in GLE with the laser parameter qualitatively. This experimental study can not only give a better understanding of the pulses formation mechanisms of these different solitons in mode-locked lasers but also provide insight of GLE systems. Furthermore, the three different kinds of solitons can be emitted in only one laser makes the laser very attractive in certain applications.

Usually, DM solitons are generated in the cavity with slightly positive dispersion. However, they can also be generated with large net positive dispersion theoretically [10]. The net positive dispersion can be increased resulting in decreasing the spectrum width [9, 11]. Previously, the slightly net positive dispersion is adopted to generate sub-100 fs pulses [12]. DM solitons with large positive dispersion are rarely studied. It's well known that ANDi exists in all normal dispersion cavities, and it can also be found in dispersion-managed cavities with large net positive dispersion both by theory [13] and experiments [8, 14], which can be explained by stability of ANDi under anomalous dispersion perturbations[13]. As a result, DM solitons and ANDi may coexist in a cavity with large net positive dispersion. GLE can be found elsewhere, and the coefficient of GLE is related to the parameter of mode-locked lasers recently [15]. It's shown as below:

$$i\psi_z + \frac{D}{2}\psi_{tt} + \psi|\psi|^2 = i\delta\psi + i\beta\psi_{tt} + i\varepsilon\psi|\psi|^2 + i\mu\psi|\psi|^4 \quad (1)$$

where $\psi$ is the electric field envelope circulating in the cavity, $z$ and $t$ denote the distance and retarded time. $D$ is the net cavity dispersion, which is negative for normal dispersion and positive for anomalous dispersion.

$$\beta = \frac{g_o}{\omega_g^2 |\beta_2|} \quad (2)$$

$$\delta = Lg_o + \ln|\theta\cos 2\alpha_3 \cos\alpha| \quad (3)$$

$$\varepsilon = -\frac{1}{3}\sin 2\alpha_2 \tan\alpha, \mu = -\frac{1}{9}\frac{\sin^2 2\alpha_3}{2\cos^2\alpha} \quad (4)$$

$$\alpha = 2\alpha_2 - \alpha_1 - \alpha_3 \quad (5)$$

where $\beta_2$ is the second order group velocity dispersion, $g_o$ is the linear gain, $\alpha_1, \alpha_2, \alpha_3$ are denoted by the angles between one eigenaxis of the plates 1,2,3 and the passing axis of the polarizer [15]. In equations (2) and (3), $\beta$ denotes gain dispersion and $\delta$ is net gain, both of them depend on the linear gain. In experiments the linear gain can be varied by changing the saturation energy of the gain through adjusting the pump power [16], which results in tuning $\beta$ and $\delta$ . $\varepsilon$ and $\mu$ represent cubic and quintic absorption terms respectively. $\beta$ , $\delta$ , $\varepsilon$ ,and $\mu$ represents the dissipative terms [16], and as shown by Q. Quraishi *et al* [17, 18]. They can be neglected in GLE when the light field is weak, and GLE is reduced to perturbed nonlinear Schrödinger equation (PNLS) which governs the dynamics of DM solitons [17]. The main formation mechanisms of DM solitons are nonlinearity and dispersion. On the other hand, when the light field is strong enough i.e. under high pump power, dissipative terms can't be viewed as small perturbations any more as they increase with pump power increment. As a result, the solution of GLE transmits from DM solitons to ANDi whose formation mechanisms are based on nonlinearity, dispersion and dissipative terms. Thus, there is a pump power threshold for switching from DM solitons to ANDi, which can be determined by experiment. Numerical study also implies that DM solitons can be switched to other solitons by gain adjustment [5].

Similaritons can be generated by putting an initial pulse of arbitrary intensity profile in an amplifier with normal dispersion [19]. However, similariton generation becomes a challenge in a laser, because the evolution of their features is monotonic, which can't be self-consistent in a laser cavity. New mechanism to reverse the evolution must be added. A filter can compensate the evolution, which has been demonstrated experimentally [3, 20]. Nonlinear polarization rotation (NPR) based on fiber components is not only a mode-locker but also a spectral filter with tunable bandwidth. The bandwidth induced by the birefringence of the fiber can be changed continuously through polarization controllers (PC) tuning [21, 22]. It has been confirmed numerically [6] that ANDi can be switched to similariton as long as the spectrum width of ANDi can be decreased to threshold value by a filter. Physically, the filter can make the evolution of similariton self-consistent. Thus, similariton generation is possible. This kind of similariton are classified as amplifier similaritons [23].

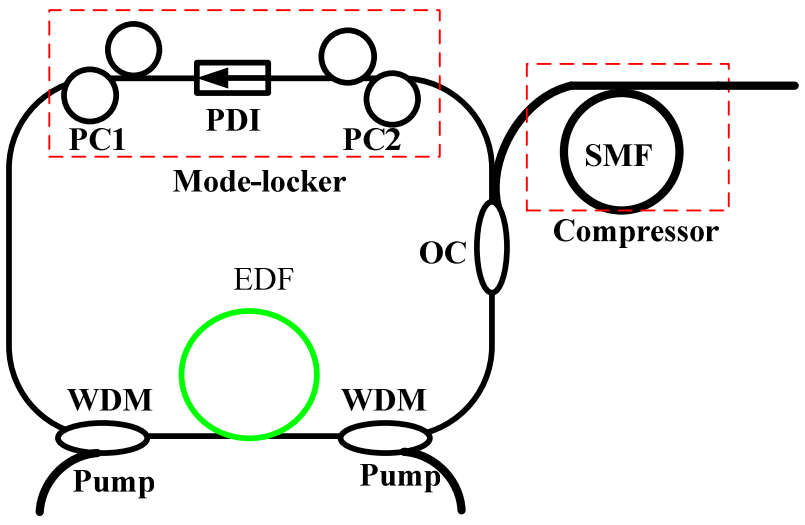

FIG. 1 Experimental configuration of the laser.

To meet these requirements, we construct a fiber laser consisting of the segments of anomalous and normal dispersion, but with large net positive dispersion. As shown in Figure 1, the laser cavity is made of a 250 cm EDF with GVD of -51 ps/ (nm·km) at 1550 nm (80 dB/m absorption ratio at 1530 nm), which is bidirectionally pumped by two 976 nm laser diodes through two wavelength division multiplexers (WDMs). The 10:90 optical coupler (OC) is used to output the laser signal. For reducing anomalous dispersion, the OC is made of dispersion-shifted fiber (DSF), so is the fiber wrapped around the two PCs. The two WDMs are made of Nufern 980 fiber. The GVD parameters of the fibers are 7 (DSF), and 4.5 (Nufern 980 fiber) ps/(nm·km) at 1550 nm, respectively. The polarization dependent isolator (PDI) is pigtailed by SMF with GVD of 17 ps/(nm·km) at 1550 nm. However, the length of SMF is minimized to only 15 cm, and thus the anomalous dispersion of PDI can be neglected. The net dispersion is 0.12 $ps^2$. In principle, this dispersion parameter can meet the requirement to generate different kinds of pulse solutions of GLE in one mode-locked laser.

Mode-locking was initiated by NPR, which relies on intensity dependent rotation of an elliptical polarization state in a length of optical fiber. With proper settings of the initial polarization ellipse and phase bias, pulse shortening occurs. We studied the laser features under weak light field at first. Gaussian type spectrum is observed when the laser is mode-locked at the pump threshold of 100 mw. Fig. 2 is the observed spectrum with 9 nm width and its Gaussian fitting (red). It's well known that the spectrum width decreases with net dispersion increment [9, 11]. The spectrum width (9 nm) is much narrower than that of typical DM solitons (60 nm) [12] as the net dispersion in our laser is nearly an order magnitude larger than DM solitons lasers. The inset of Fig. 2 is the autocorrelation trace, and the pulse is 355-fs assuming a Gaussian shape. DM solitons can be distinguished from other solitons by Gaussian type spectrum [9]. In Fig. 2, the spectrum is fitted by the Gaussian function quite well indicating that it's DM solitons. Furthermore, the time-bandwidth product is 0.4 closed to 0.44 for Gaussian pulses. H. A. Haus theoretically showed that the spectrum width of DM solitons decreases with gain increment [10], but no experiment has confirmed that up to date. We experimentally study this relationship to demonstrate this characteristic of DM solitons. The pump power is continuously increased with an interval of 3 mw. Remarkably, as shown in Fig. 3, the spectrum width is decreased continuously, and the shape is still Gaussian type. The width decreases from 13 nm to 2 nm when pump power is tuning from 100 mw to 315 mw as seen in Fig. 4. Further increasing the pump power will destroy mode-locking state, and continuous wave appears in the spectrum as a spike as shown in Fig. 5. Breaking down of DM solitons means that the dissipative terms of GLE can't be viewed as small perturbations any more under strong light field, resulting in invalid of DM soliton solutions of GLE.

As mentioned above, there is a pump power threshold for switching from DM solitons to ANDi. We find experimentally that this threshold is just the one when DM solitons are destroyed at the pump power of 319 mw. Adjusting PC can generate another nonlinear wave after DM solitons are broken. Fig. 6 shows the spectrum of the observed new nonlinear wave, which is steep at edges and deep in the top. Such a spectrum is a typical characteristic one of ANDi [8, 24]. The pump power of 319 mw is the threshold of the transition between DM solitons and ANDi. It's the critical value that makes the governing equation of the laser system changes from PNLS to GLE. In other words, the related coefficients of GLE can be neglected as small perturbations when the pump power is below 319 mw. They can't be viewed as small perturbations and play important roles in generating ANDi when the pump power is above 319 mw. The inset of Fig. 6 shows the autocorrelation trace of observed ANDi pulses. The pulse width is 192-fs after compressed by SMF.

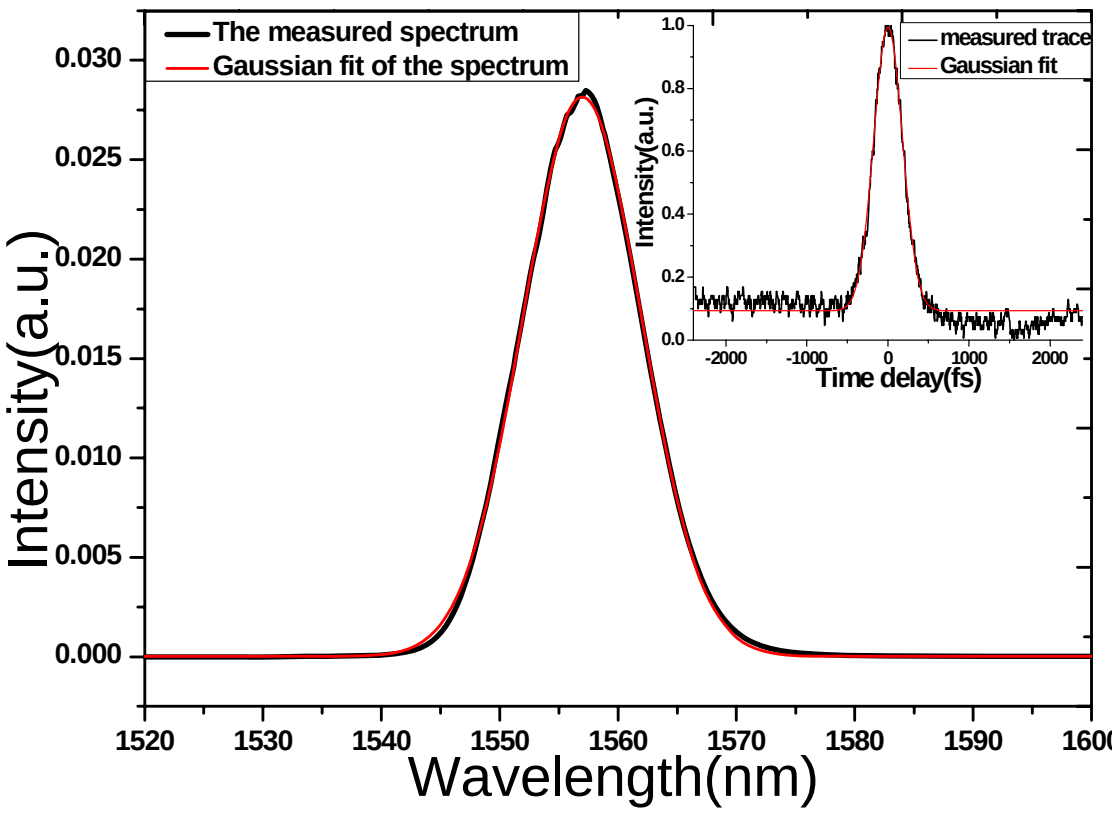

FIG. 2 The spectrum and autocorrelation trace of (inset) the DM solitons (black) and Gaussian fit (red).

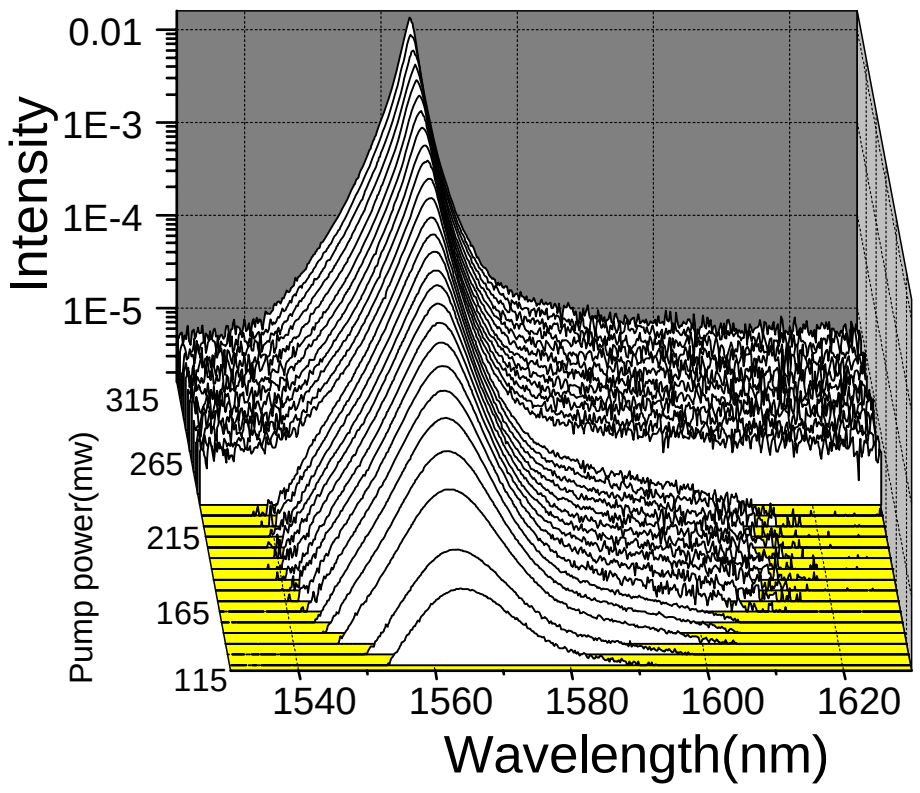

FIG. 3 The spectrum scaling of DM solitons vs pump power.

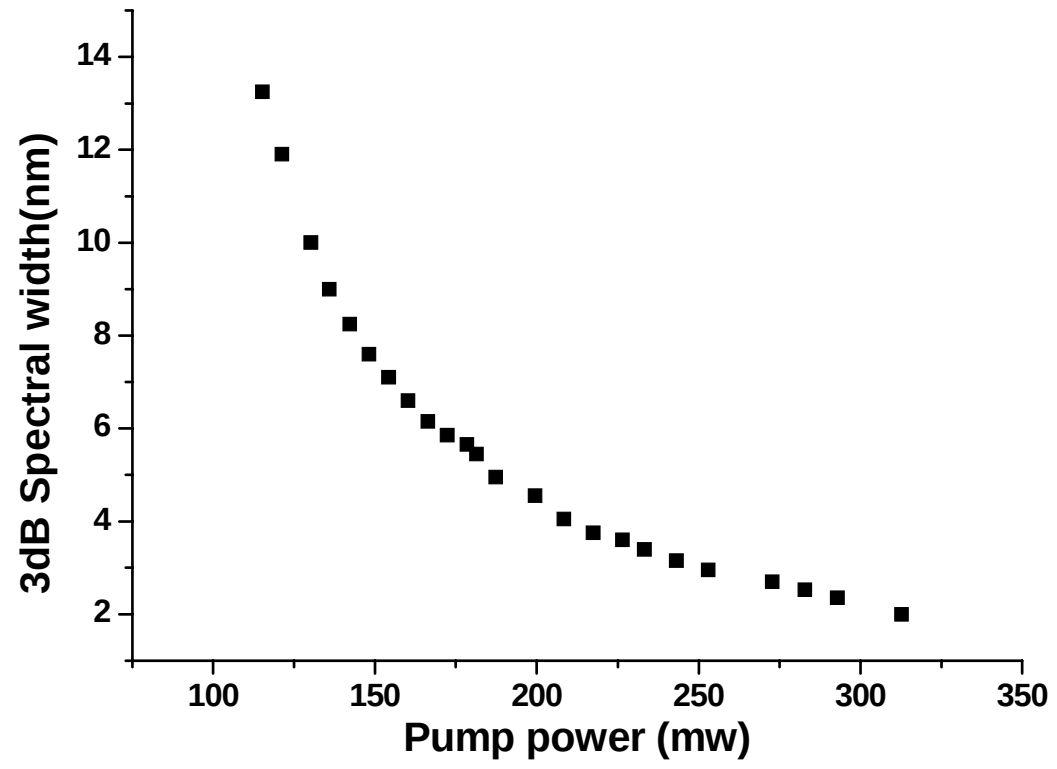

FIG. 4 The spectrum width scaling vs pump power.

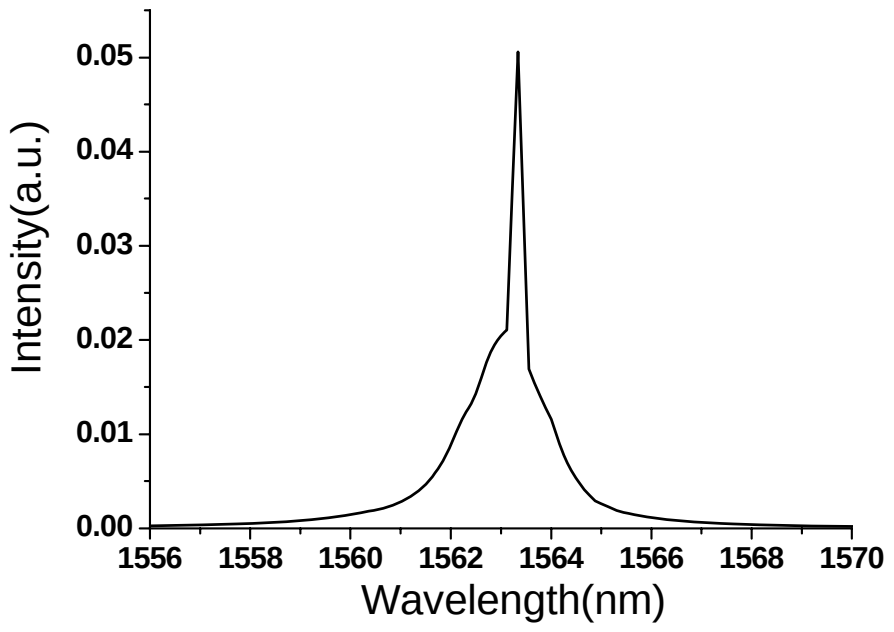

FIG. 5 The spectrum of cw generation.

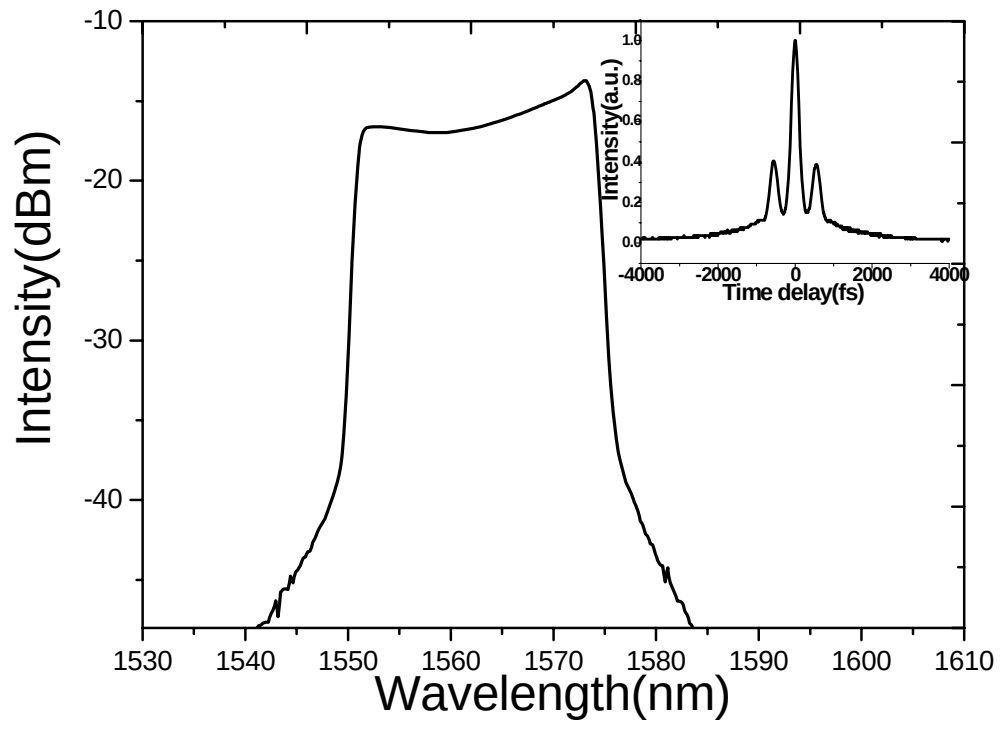

FIG. 6 The spectrum and autocorrelation trace (inset) of ANDi.

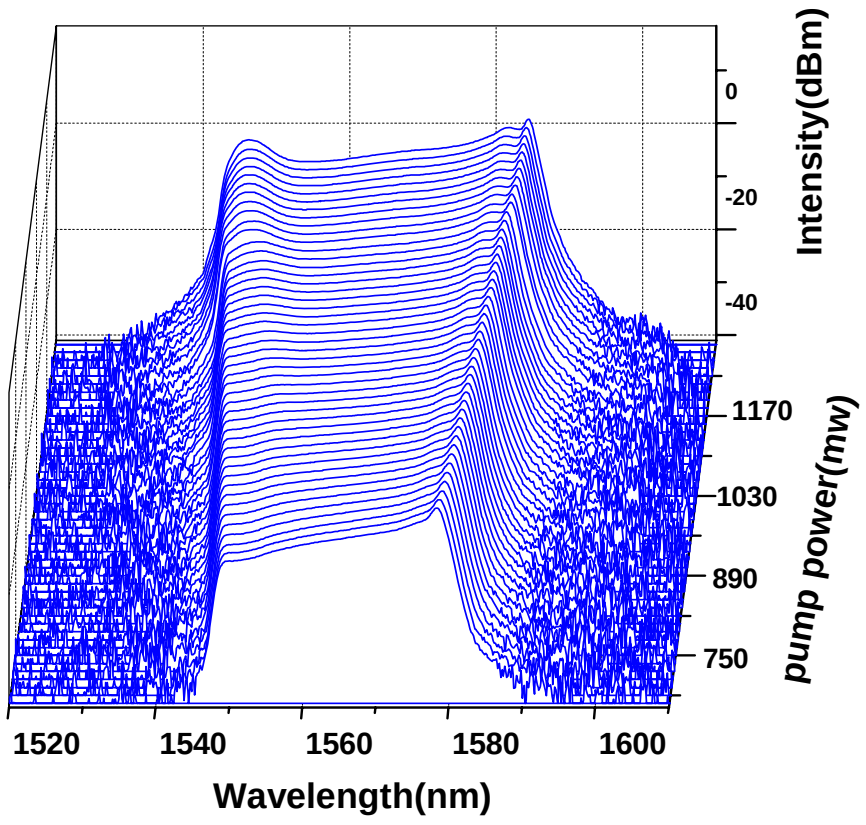

FIG. 7 Pulses spectrum of ANDi scaling with pump power.

Kalashnikov [7]and Cabasse [8] theoretically presented that the spectrum width of ANDi increases with the pump power increment. As shown in Fig. 7, we experimentally examine this conclusion. The width increases with pump power increment without changing shape, which is opposite to DM solitons. Comparing these different features of ANDi and DM solitons, we get qualitative understanding about them. The light field gets stronger as the pump power increases, which causes DM solitons to switch to ANDi. However, due to the stability of DM solitons, they can be broadened in time domain to decrease the strength of the light field through narrowing the spectrum width. Thus, the spectrum width of DM solitons decreases with the pump power increment. The minimum spectrum width of ANDi is about 25 nm which is much wider than DM solitons (13 nm maximum) in the laser. This indicates that gain dispersion becomes more obvious than DM solitons in the pulse shaping mechanism [8] considering the gain bandwidth of EDF (30 nm). Gain dispersion can't be viewed as small perturbations and play important roles in ANDi generation.

NPR based on fiber component is also a spectral filter with tunable bandwidth. Slightly tuning one PC paddle clockwisely (counter-clockwisely), the spectral width increases (decreases) without changing in shape when ANDi is generated. Remarkably, when the spectrum width of ANDi decreases to certain value (~ 30 nm), the spectrum shape suddenly changes, which is shown in Fig. 8. Compared to the one of ANDi, the dip in the top is raised and the steep edges become smooth. It may be amplifier similariton after considering its formation mechanism. We examine the feature by fitting the spectrum with parabolic function. In Fig. 8, the spectrum is well fitted by the parabolic function (red). The parabolic spectrum is the distinguished shape of similaritons[25]. Gaussian fitting (green) is also present for comparison. The spectrum deviates from the Gaussian fitting indicates that it's not DM soliton. We can thus claim that the pulse is indeed the similariton. Fig. 9 shows the autocorrelation trace, and the pulse width is 197-fs. It should be noted that the gain in amplifier also plays an important role in similariton generation [26]. Decreasing the spectral width of ANDi to the minimum by PC tuning won't generate similaritons if the total pump power is lower than a certain threshold value (backward pump 49 mw, forward pump 546 mw). However, then increasing the pump power above the threshold, ANDi changes to similaritons without PC adjustment. Thus, we experimentally demonstrate that both the gain of amplifier and the bandwidth of spectral filter play important roles in similariton generation, which are consistent with the theoretical studies [26, 27].

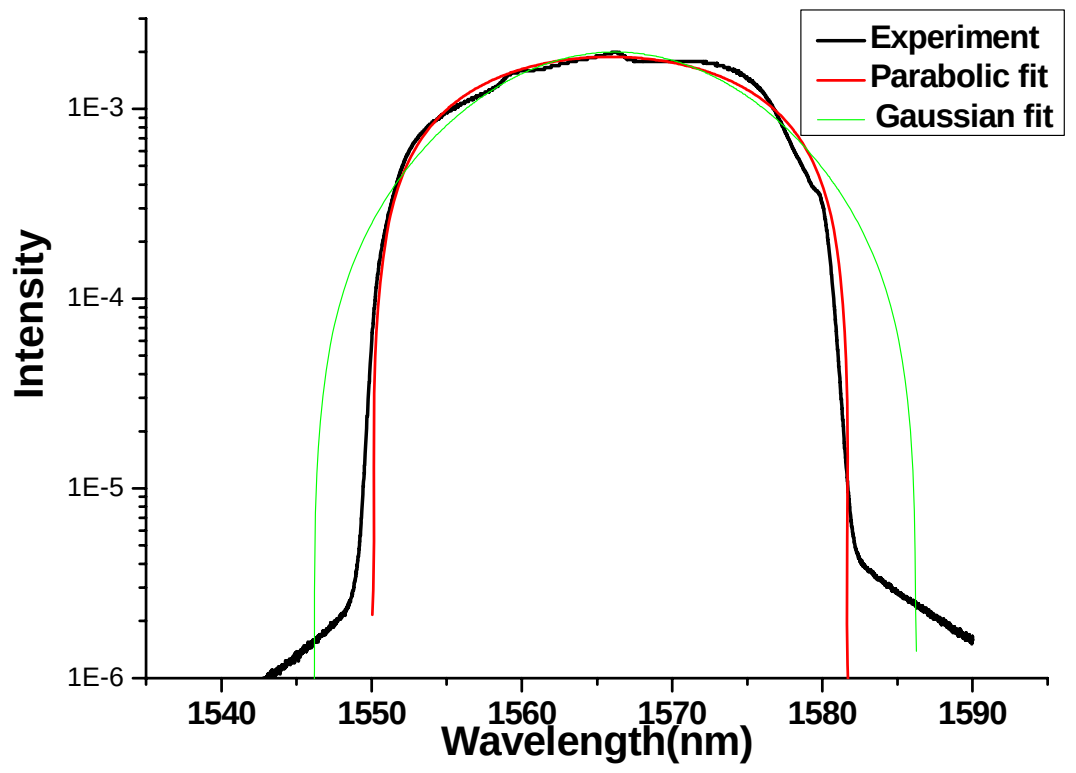

FIG. 8 The spectrum of similaritons.

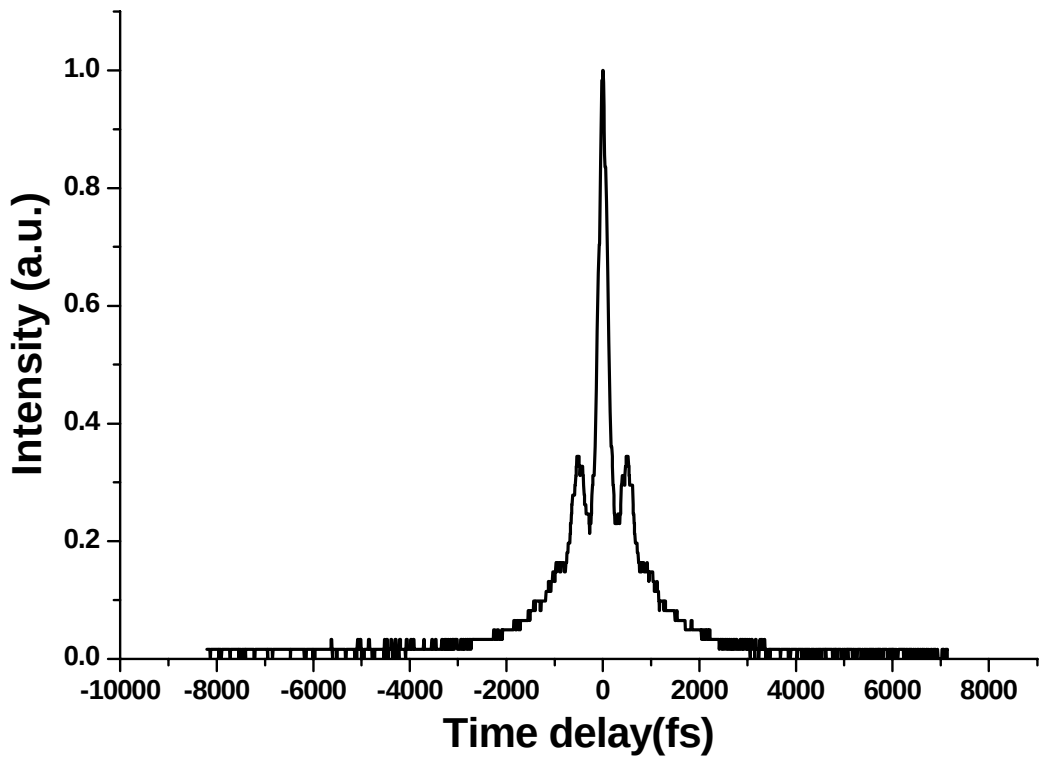

FIG. 9 The autocorrelation trace of similaritons.

Finally, three different kinds of solutions of GLE are observed in a laser with net positive dispersion. Mathematically, the solutions depend on the coefficients of GLE, and varying the coefficients can obtain the solutions respectively which correspond to different kinds of nonlinear waves physically. DM solitons governed by PNLS are formed in Hamilton systems, which are generated by balance between nonlinearity and dispersion. ANDi and similaritons are dissipative solitons generated in dissipative systems, and in addition to nonlinearity and dispersion, dissipative effects such as gain dispersion, spectral filtering also account for their generation. ANDi can be shaped to similaritons once the evolution of similaritons can be self-consistent by a spectral filter in the laser. Transitions between different solitons indicate that Hamilton systems and dissipative systems can be switched to each other in a mode-locked laser, which indicates that the two systems are related to each other. Transitions between different solitons in a single laser reveal interesting features of GLE, which were only theoretically investigated before. We only monitor the spectrum features of the pulses in the experiment. The counterpart in time domain is hard to be measured due to its ultrashort duration. However, thanks to different spectrum shape of the three solitons, spectrum measurement can distinguish them effectively. Mode-locked fiber lasers are excellent tools to study nonlinear physics. Dispersion, gain dispersion, spectral filtering and other effects exist simultaneously in mode-locked fiber lasers, and our study shows that they can be changed in mode-locked lasers through pump power and the birefringence of fiber, which contribute to transitions between different solitons.

In conclusion, we observe three different kinds of solitons in a mode-locked fiber laser. Transitions between them are realized by gain adjustment and the birefringence effect. The three kinds of pulses show different characteristics especially the spectrum shape. This study not only gives insight of GLE but also shows the formation mechanisms of different solitons much more clearly in mode-locked fiber lasers. Furthermore, due to different properties of the three solitons, transitions between them in one laser are inherently attractive.

The authors acknowledge the support from the National Natural Science Foundation of China (Grant 10874118), and the key project of the Ministry of Education of China (Grant 109061).

*Corresponding author: lizhan@sjtu.edu.cn